\documentclass{AIAA}

\usepackage{amsmath}
\usepackage{amssymb}
\usepackage{graphicx}
\usepackage{xcolor}
\usepackage{subfigure}

\begin{document}

\title{Low-energy capture of asteroids onto KAM tori}

\author{Patricia E. Verrier\footnote{Research Associate, Advanced Space Concepts Laboratory, Department of
Mechanical and Aerospace Engineering.} and Colin R. McInnes\footnote{Director, Advanced Space Concepts Laboratory, Department of
Mechanical and Aerospace Engineering. Currently: James Watt Chair, Professor of Engineering Science, School of Engineering, University of Glasgow; colin.mcinnes@glasgow.ac.uk.}}
\affiliation{University of Strathclyde, Glasgow, Scotland G1 1XJ, United Kingdom}

\maketitle

\section{Introduction}

Engineering the artificial capture of asteroids in the neighbourhood of the Earth is a challenging problem. The simplest method would be to alter an asteroid's Jacobi constant so that it is trapped within the zero velocity curves (ZVCs) in the neighbourhood of the Earth. This has been investigated by \cite{Baoyin2010} who considered a range of NEOs in the context of the Earth-Sun Circular Restricted Three-Body Problem (CRTBP) and determined the $\Delta v$ needed to close the zero velocity curves and truly capture the object. The lowest $\Delta v$ required was found to be of the order of 400 $\mathrm{m\ s^{-1}}$. Another method suggested by \cite{Hasnain2012} uses a variant of patched conics to transfer an asteroid into the Earth's Hill sphere combined with an impulsive transfer onto a bound geocentric orbit.  They identify 23 asteroids that could be captured within 10 years, which have final capture $\Delta v$ requirements ranging from 700 $\mathrm{m\ s^{-1}}$ to 4900 $\mathrm{m\ s^{-1}}$. Similarly, \cite{Sanchez2011} and \cite{Sanchez2012a} consider various capture strategies involving patched conics to create resource maps of accessible NEOs within a given $\Delta v$ range. 

An alternate method for capturing asteroids has been suggested by \cite{Yarnoz2013} who find a new class of `Easily Retrivable Objects' (EROs) among the near-Earth object (NEO) population. These objects are those that can be easily transferred onto the invariant manifolds associated with Lyapunov and Halo orbits around the Earth-Sun $L_1$ and $L_2$ points. They find a list of 12 objects that can be transferred into the Earth locale with a $\Delta v$ of less than 500 $\mathrm{m\ s^{-1}}$. The problem of finding long term stable parking orbits and associated transfers for such asteroids has received less attention though. In \cite{Verrier2014}, families of linearly stable co-orbital periodic orbits about a central body are identified. Such orbits are potential locations for `parking' asteroids about an object such as the Moon, although transfer strategies to such orbits are not immediately apparent.

A different approach to the problem is used by \cite{Urrutxua2014}, who consider `temporarily captured asteroids' (TCAs) such as 2006 $\mathrm{RH}_{120}$. TCAs are objects that temporarily come within the Hill sphere of the Earth, and remain for a short period of time (months to years). Using a dynamical model of the Earth-Moon-Sun system they find that a relatively low $\Delta v$ of 32 $\mathrm{m\ s^{-1}}$ applied gradually over several months can extend the capture phase by several years.  

Another method is introduced by \cite{Borum2012}, who considers the result of close approaches of binary asteroids with the Earth, demonstrating that under some conditions a binary exchange can occur. The result of such an encounter is the capture of one of the asteroids about the Earth, while the other is ejected from the system on an escape trajectory.

In developing these new strategies for asteroid capture, astronomical theories explaining the capture of irregular satellites of the giant planets are relevant. There are three main theories: gas drag (capture due to dissipation induced by the gas drag in the early solar system), pull down capture (capture as a protoplanet's Hill sphere expands as its mass increases) and the broad category of $n$-body interactions. Of the first two mechanisms pull down capture does not readily translate to methods for engineering asteroid capture, while ideas relating to gas drag have been considered by \cite{Sanchez2012b} who propose a method of aero-assisted capture that exploits the dissipation of energy due to atmospheric drag to capture asteroids Earth-bound orbits. The last category presents further opportunities, and has been the focus of several studies already, such as the binary exchange mechanism proposed by \cite{Borum2012} or the invariant manifold method of \cite{Yarnoz2013} both discussed above. 

A review of the various n-body interaction theories in the context of formation of irregular satellites is presented in \cite{Philpott2010}. These fall into the broad categories of collisions, n-body interactions due to resonance crossings of the giant planets and exchange reactions, such as the disruption of a binary asteroid. The disruption of a binary asteroid has been investigated in the context of asteroid capture by \cite{Borum2012}, but presents engineering challenges as it is very sensitive to the configuration of the asteroid during the encounter. 

Another dynamical mechanism for the capture of irregular satellites is chaos-assisted capture, as proposed by \cite{Astakhov2003}. Here the phase space of the Sun-Jupiter CRTBP is shown to consist of islands of Kolmogorov-Arnold-Moser (KAM) orbits separated from scattering orbits by a thin chaotic layer. A small perturbation such as dissipative gas drag can then capture a chaotic object into this regular region of phase space. In a two-dimensional model the captured object can then not escape, while in three dimensions Arnold diffusion means it can move out through the KAM tori exponentially slowly, and is thus effectively captured on astronomical timescales. They show that this mechanism is viable in the context of both the two-dimensional and three-dimensional Sun-Jupiter CRTBP. In \cite{Astakhov2004}, the chaos-assisted capture mechanism in the Eccentric Restricted Three-Body Problem (ERTBP) is shown to be robust to perturbations from other planets as well. In \cite{Astakhov2005} the formation of Kuiper-belt binaries using this capture mechanism is studied in the context of the three- and four-body Hill problem. In this case a temporary chaotic binary is moved to a regular trajectory through scattering encounters with other objects. The chaos-assisted capture mechanism provides an interesting possibility for engineering a relatively low energy artificial capture of an asteroid in the Earth's neighbourhood. 

In this work we show how a temporary captured asteroid that passes close to a regular region of phase space can be moved onto the nearby KAM tori and essentially be captured permanently in the Earth's neighbourhood, via the chaos-assisted capture mechanism of \cite{Astakhov2003}. We demonstrate the method and the relatively low $\Delta v$ required for an example asteroid trajectory.

We use the two-dimensional Hill problem as a simple first model of the Earth-Sun system, and calculate trajectories and Poincar\'e surfaces of section (SOS) using a 6th order Runge-Kutta-Nystr\"om (RKN) symplectic integration scheme. Although this does not capture the three-dimensional dynamics or the effect of the Moon on the near-Earth environment, it does demonstrate the basic principles of the method and provides a demonstration of the capture energies required. Further development is not expected to vastly alter either of these, merely the details of the trajectories. Details of the mathematical model and integration scheme are given in the next section, followed by the results for the example trajectories.

\section{Hill problem surface of section}

The Earth-Sun system can be modelled simply using the two-dimensional Hill problem. To generate a Poincar\'e surface of section (SOS) numerical methods are needed to integrate the equations of motion.  One such method is a symplectic integration scheme. Such integrators preserve the phase space structure and so are advantageous for the generation of Poincar\'e SOSs. A 6th-order symplectic integration scheme for the Hill problem can be easily constructed by extending the method of \cite{Quinn2010}, who derive a second-order symplectic scheme for the planar Hill problem, to a higher order scheme such as given by \cite{McLachlan1995}. The planar Hill problem has Hamiltonian 
\begin{equation}
H(x, y, p_x, p_y) = \frac{1}{2}(p_x^2 + p_y^2) + (y p_x -x p_y) - x^2 + \frac{1}{2} y^2 -\frac{1}{(x^2+y^2)^{1/2}}
\end{equation}
where $x$ and $y$ are the scaled coordinates relative to the Earth in the rotating frame and $p_x=\dot{x}-y$ and $p_y=\dot{y}+x$ are the corresponding conjugate momenta. This is the dimensionless formulation of the Hill problem and uses units such that the mean motion of the Earth about the Sun is unity. It can be derived from the CRTBP by moving the origin to the location of the Earth and taking the limit of the mass ratio going to zero. In the Hill problem the energy $E$ is a constant of the motion and is often used in the form of the Hill-Jacobi constant $C_H = -2E$. In \cite{Quinn2010}, a canonical transform is introduced that simplifies the Hamiltonian and allows it to be split into two analytically integrable parts. The new coordinates and momenta are given by $X = x$, $Y = y$, $P_X = p_x + y$, $P_Y = p_y + x$ and the transformed Hamiltonian is
\begin{equation}
H(X, Y, P_X, P_Y) = \frac{1}{2} (P_X^2 + P_Y^2) - 2 X P_Y + \frac{X^2}{2} - \frac{1}{(X^2+Y^2)^{1/2}}.
\end{equation}
This Hamiltonian can be split as $H = H_1 + H_2$ with
\begin{eqnarray}
H_1 &=& \frac{1}{2} (P_X^2 + P_Y^2) - 2 X P_Y \\
H_2 &=& \frac{X^2}{2} - \frac{1}{(X^2+Y^2)^{1/2}}.
\end{eqnarray}
The equations of motion associated with both $H_1$ and $H_2$ are now trivially integrable. As given by \cite{Quinn2010} the evolution from time $t$ to $t+\tau$ under $H_1$ is given by
\begin{eqnarray}
X(t + \tau)   &=& X(t) + \tau P_X(t) + \tau^2 P_Y(t)\nonumber\\
Y(t + \tau)   &=& Y(t) + \tau \big(P_Y(t) - 2X(t) \big) - \tau^2 P_X(t) - \frac{2}{3} \tau^3 P_Y(t) \nonumber\\
P_X(t + \tau) &=& P_X(t) + 2 \tau P_Y(t)\nonumber\\
P_Y(t + \tau) &=& P_Y(t)
\end{eqnarray}
and the evolution from time $t$ to $t+\tau$ under $H_2$ by
\begin{eqnarray}
X(t + \tau)   &=& X(t) \nonumber\\
Y(t + \tau)   &=& Y(t) \nonumber\\
P_X(t + \tau) &=& P_X(t) - \tau \left (X(t) + \frac{X(t)}{(X(t)^2+Y(t)^2)^{3/2}} \right)\nonumber\\
P_Y(t + \tau) &=& P_Y(t) - \tau \left (\frac{Y(t)}{(X(t)^2+Y(t)^2)^{3/2}}\right).
\end{eqnarray}
In \cite{Quinn2010}, the authors further partition the term $H_1$ into two pieces to permit quick detection of collisions between particles for simulations of ring dynamics. However, this is not required here and the splitting above is sufficient. Further, this splitting $H=H_1(P,X)+H_2(X)$ satisfies the condition to use a Runge-Kutta-Nystr\"om (RKN) symplectic method, as the term $H_1$ is quadratic in the momenta (see e.g. \cite{McLachlan1995}). Such a method is the 6th order RKN symplectic scheme given by \cite{McLachlan1995} (also used by \cite{Waldvogel1996} for the Hill problem in regularized coordinates). We use a 6th order method for accuracy as the two Hamiltonians $H_1$ and $H_2$ are fairly comparable in size. The RKN symplectic scheme is more accurate than other schemes at this order, but requires more force evaluations per step. However this is not an issue here as we use a symplectic algorithm for accuracy as explained above, rather than speed. The RKN scheme was found to be sufficient to generate the SOS needed for this work.

Let $\phi_{H1}$ represent the mapping associated with the Hamiltonian $H_1$ and $\phi_{H2}$ that with $H_2$. The RKN-6 scheme approximating the evolution under the Hamiltonian $H=H_1+H_2$ over timestep $\tau$ is
\begin{eqnarray}
\phi_{RKN6} (\tau) &=& \phi_{H_1}(a_1 \tau) \phi_{H_2} (b_1 \tau) \phi_{H_1}(a_2 \tau) \phi_{H_2} (b_2 \tau) \phi_{H_1}(a_3 \tau) \phi_{H_2} (b_3 \tau) \phi_{H_1}(a_4 \tau) \phi_{H_2} (b_4 \tau) \nonumber \\
{}& & \qquad \phi_{H_1}(a_4 \tau) \phi_{H_2} (b_3 \tau) \phi_{H_1}(a_3 \tau) \phi_{H_2} (b_2 \tau) \phi_{H_1}(a_2 \tau) \phi_{H_2} (b_1 \tau) \phi_{H_1}(a_1 \tau) \phi_{H_2} (b_1 \tau) 
\end{eqnarray}
where the coefficients $a_i$ and $b_i$ are
\begin{eqnarray}
a_1 &= -1.01308797891717472981 & b_1 = 0.00016600692650009894  \nonumber\\
a_2 &= 1.18742957373254270702  & b_2 = -0.37962421426377360608 \nonumber\\
a_3 &= -0.01833585209646059034 & b_3 = 0.68913741185181063674  \nonumber\\
a_4 &= 0.34399425728109261313  & b_4 = 0.38064159097092574080
\end{eqnarray}
as given by \cite{McLachlan1995}. 

The SOS is calculated as follows. The SOS chosen is the plane in $x$-$y$ defined by $p_x = 0$ and with $\dot{y} > 0$ (see the following Section for more details on this choice) i.e. it is a plot of the points at which trajectories cross the $x-y$ plane with zero $x$-momentum and positive $y$-velocity. 

A trajectory is generated by selecting an initial point on the section and using the constraints $p_x=0$ and $\dot{y}>0$ along with the energy integral to calculate the initial momenta of the orbit. The initial state is then propagated forwards in time using the symplectic integration scheme described above. Intersections of the trajectory with the SOS are checked for after each timestep by looking for a change in sign of $p_x$ to detect the crossing of $p_x=0$. If a crossing is detected the exact location of the intersection is then interpolated. With a small enough timestep this provides an accurate calculation of the SOS. The integration length was chosen to be sufficient to fully populate the KAM tori. For the results presented here a timestep of 0.005 in dimensionless units is used (approximately 6 hours) and the integration length is approximately 0.8 MYr. The method was tested by computing specific sections that could be compared to results in the literature.

The initial points for each SOS are selected using an interactive GUI that generates the trajectories from points selected with the mouse. This allows individual features on the surface to be fully investigated, which could otherwise be missed if a random set of initial points was used instead.

\section{Chaos-assisted artificial capture of an asteroid}

The SOSs of the planar Hill problem are well studied, see for example \cite{Simo2000} and \cite{Astakhov2005}. As found in these studies the phase space near the central object (the Earth here) at energies at above that of the two Lagrangian points consists of KAM regions surrounded by thin chaotic layers embedded in regions of escape trajectories. We select at random an example initial chaotic asteroid trajectory that enters the Hill sphere, passes near to a region of KAM tori surrounding a prograde periodic orbit (one of the $g^\prime$ family) and subsequently escapes the Earth neighbourhood. The trajectory is somewhat arbitrary, but sufficient to demonstrate the method and the energy requirements. The energy level of the initial asteroid orbit is $E=-2.152$ (in Hill units), which is similar to the examples given by \cite{Astakhov2005} for the Hill problem for binary asteroids. At this energy level the ZVCs are open at the Lagrangian points, but the Hill sphere is still a distinct region separated from the rest of the phase space by excluded areas. 

\begin{figure}
\subfigure[\label{fig1}The initial trajectory and ZVC at $E=-2.1520$]{
\includegraphics[width=0.45\textwidth]{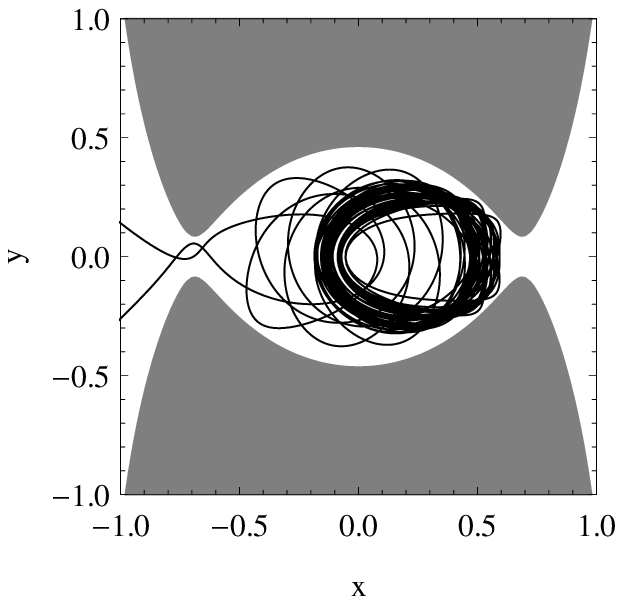}}
\subfigure[\label{fig4}The new trajectory at $E=-2.1535$]{
\includegraphics[width=0.45\textwidth]{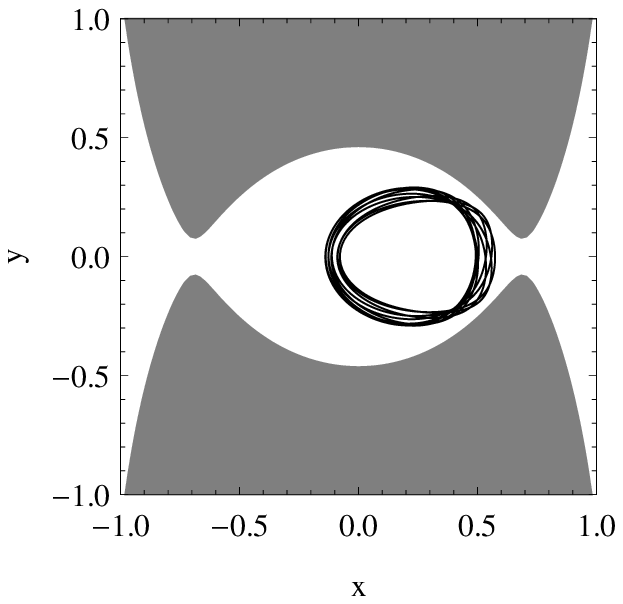}}
\caption{The initial and final example trajectories in configuration space. The corresponding ZVC surface is also shown in gray in each case (the ZVC closes at $E \approx −2.1635$)}
\end{figure}

\begin{figure}
\includegraphics[width=\textwidth]{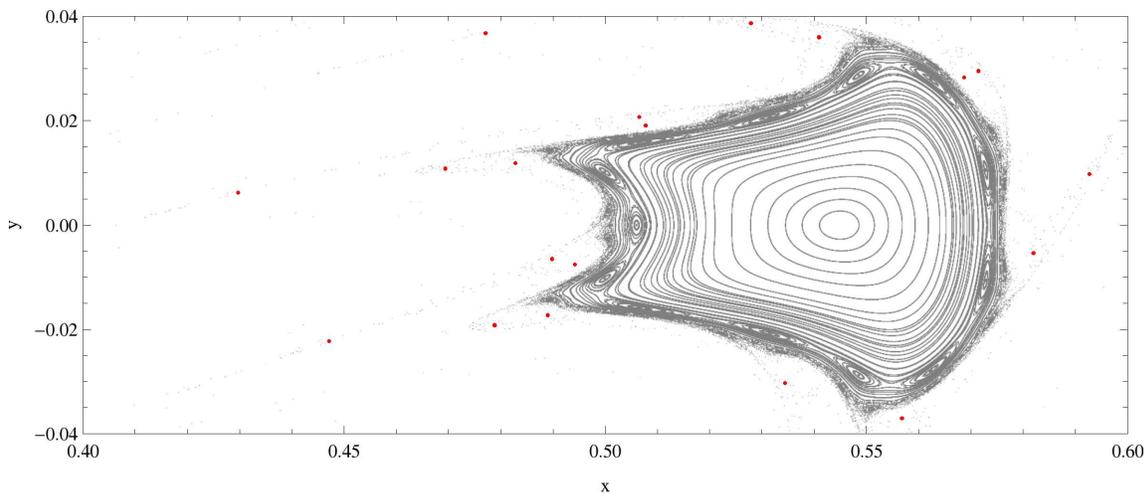}
\caption{\label{fig2} The surface of section for the initial trajectories energy level $E=-2.152$. The crossings of the example trajectory are plotted in red}
\end{figure}

The example initial trajectory is shown in configuration space in Figure \ref{fig1}, with the corresponding surface of section (SOS) in Figure \ref{fig2}. The SOS used is the plane $x$-$y$ that has $p_x = 0$ and $\dot{y} > 0$, as used by \cite{Astakhov2003, Astakhov2005} for the planar CRTBP and planar Hill problem. The SOS clearly shows the KAM region surrounded by a thin region of chaotic motion. Trajectories that pass into this region of chaos are sometimes said to be `sticky', as they stay within it for a number of crossings of the SOS, often `sticking' to tori embedded within it. The behaviour can be seen in the $x$-$y$ plot of the trajectory in Figure \ref{fig1}, the object enters the Hill sphere, roughly follows the shape of a $g^\prime$ family orbit for a number of revolutions, before exiting the sphere once more. 

\begin{figure}[!ht]
\subfigure[$E = -2.1540$]{
\includegraphics[width=0.45\textwidth]{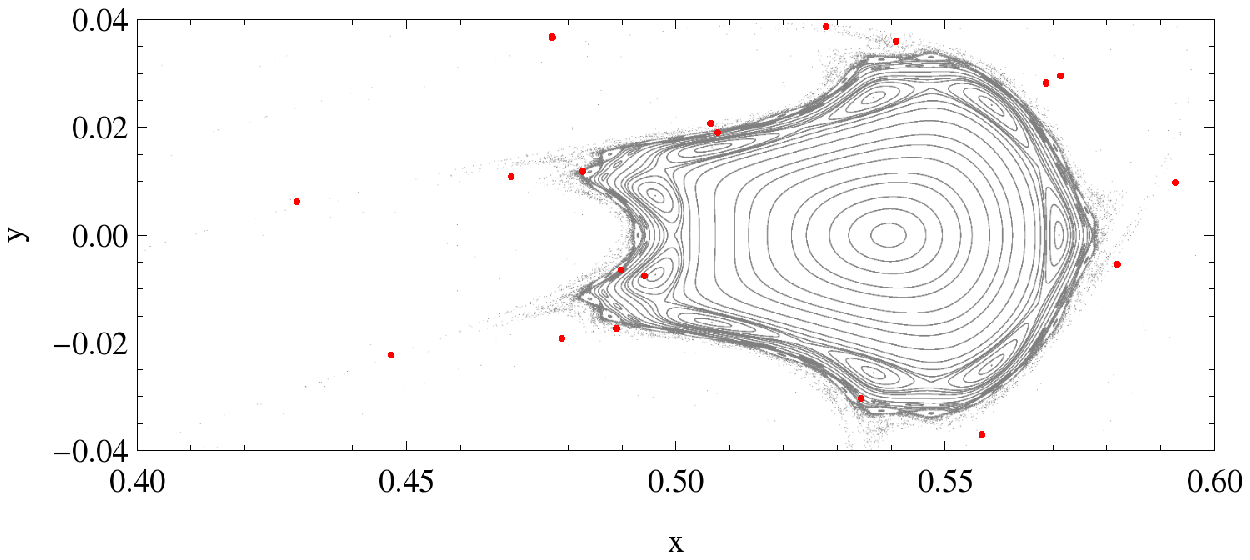}}
\subfigure[\label{figdv}$E = -2.1535$]{
\includegraphics[width=0.45\textwidth]{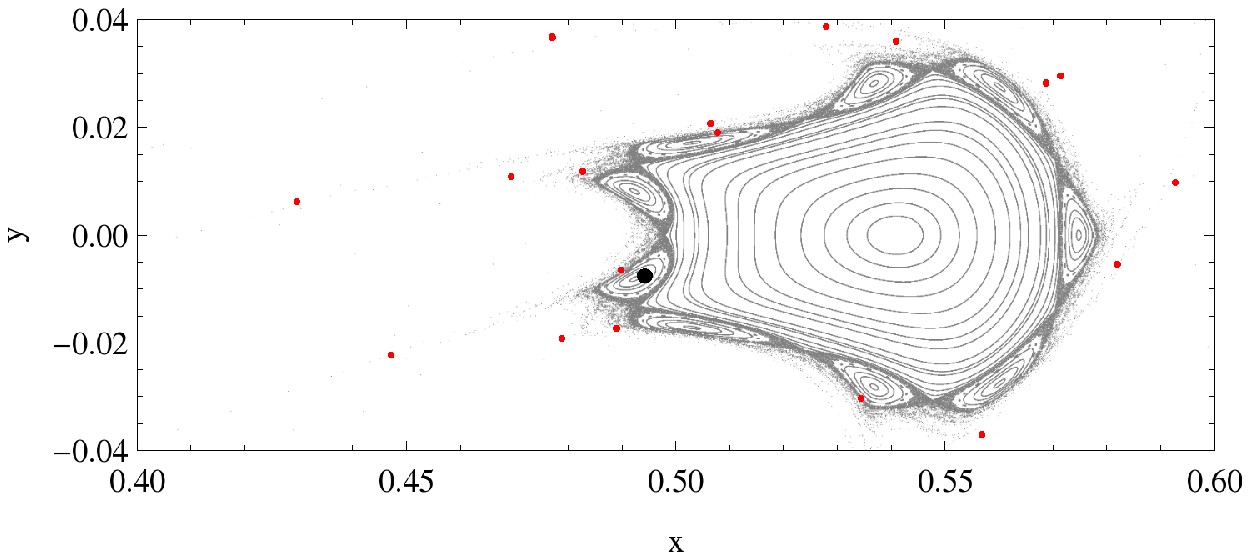}}\\
\subfigure[$E = -2.1530$]{
\includegraphics[width=0.45\textwidth]{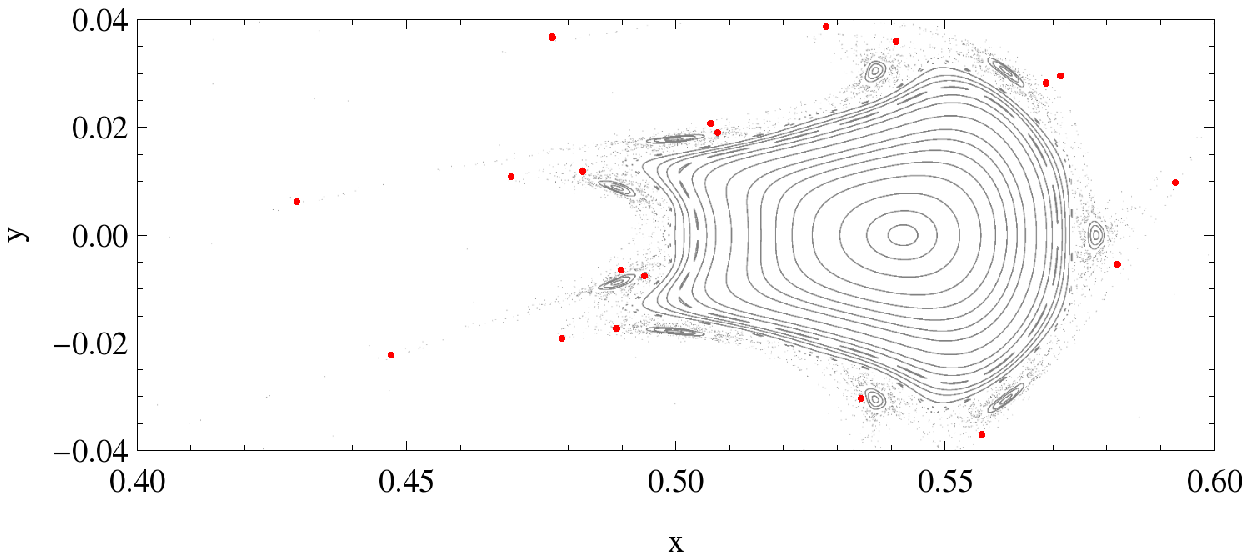}}
\subfigure[$E = -2.1525$]{
\includegraphics[width=0.45\textwidth]{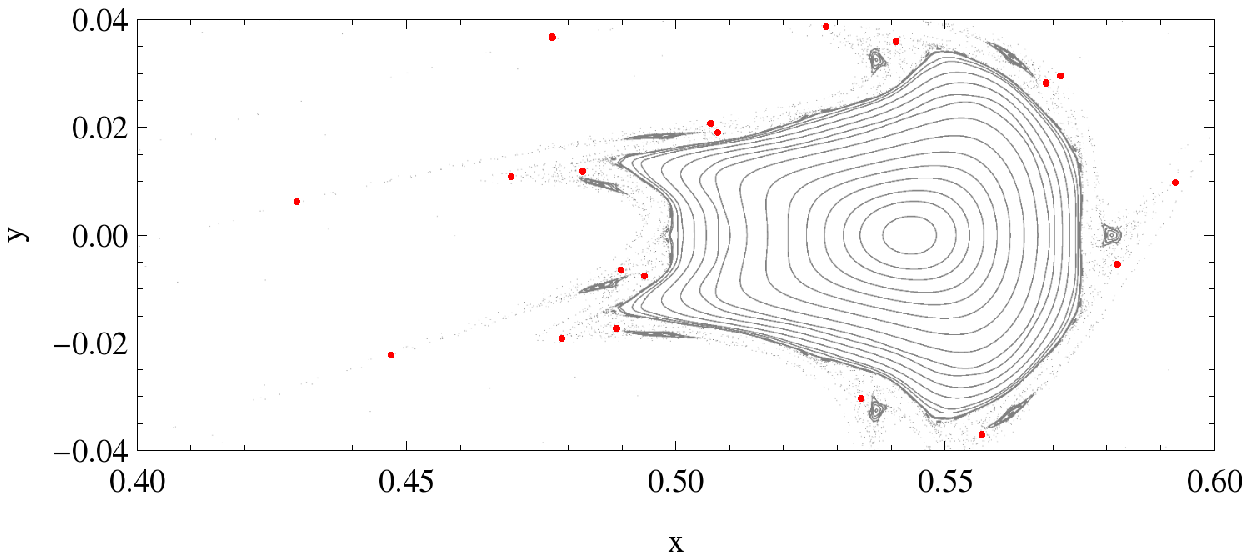}}\\
\subfigure[$E = -2.1515$]{
\includegraphics[width=0.45\textwidth]{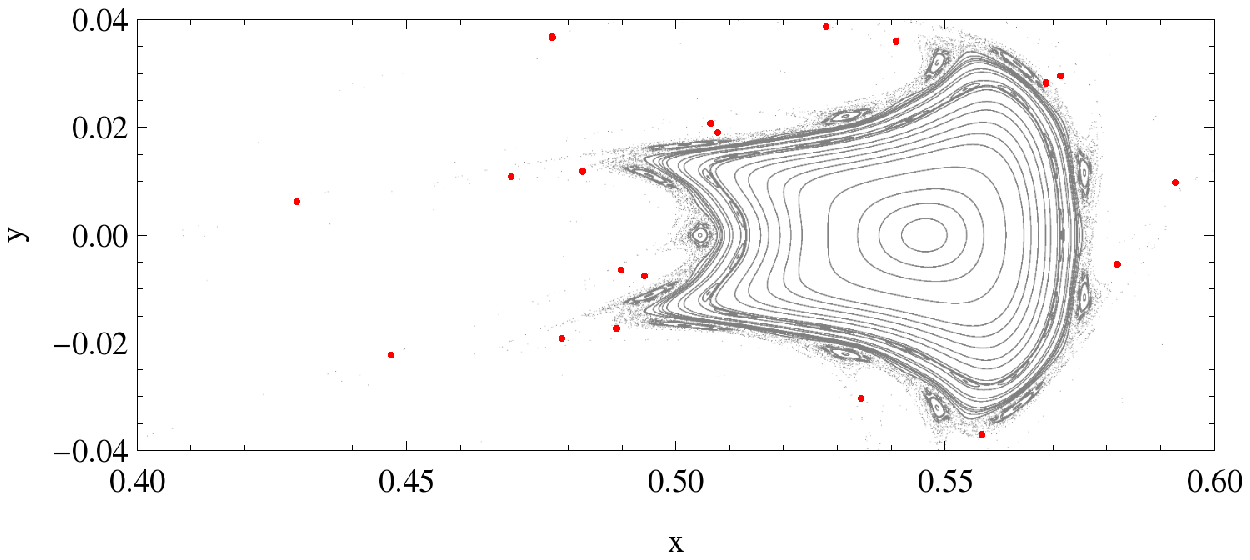}}
\subfigure[$E = -2.1510$]{
\includegraphics[width=0.45\textwidth]{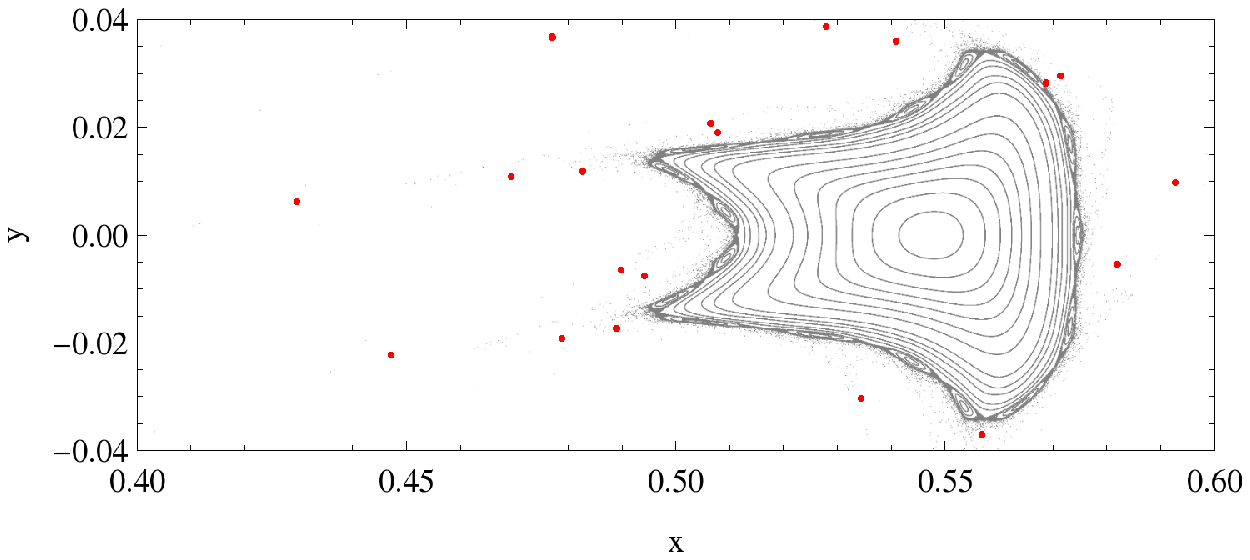}}\\
\subfigure[$E = -2.1505$]{
\includegraphics[width=0.45\textwidth]{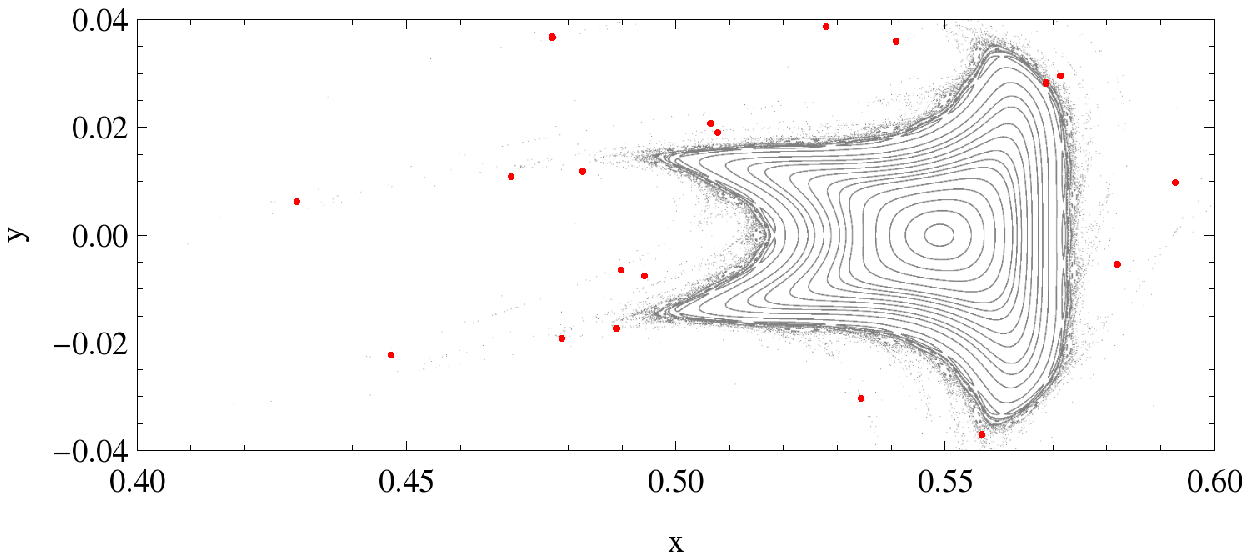}}
\subfigure[$E = -2.1500$]{
\includegraphics[width=0.45\textwidth]{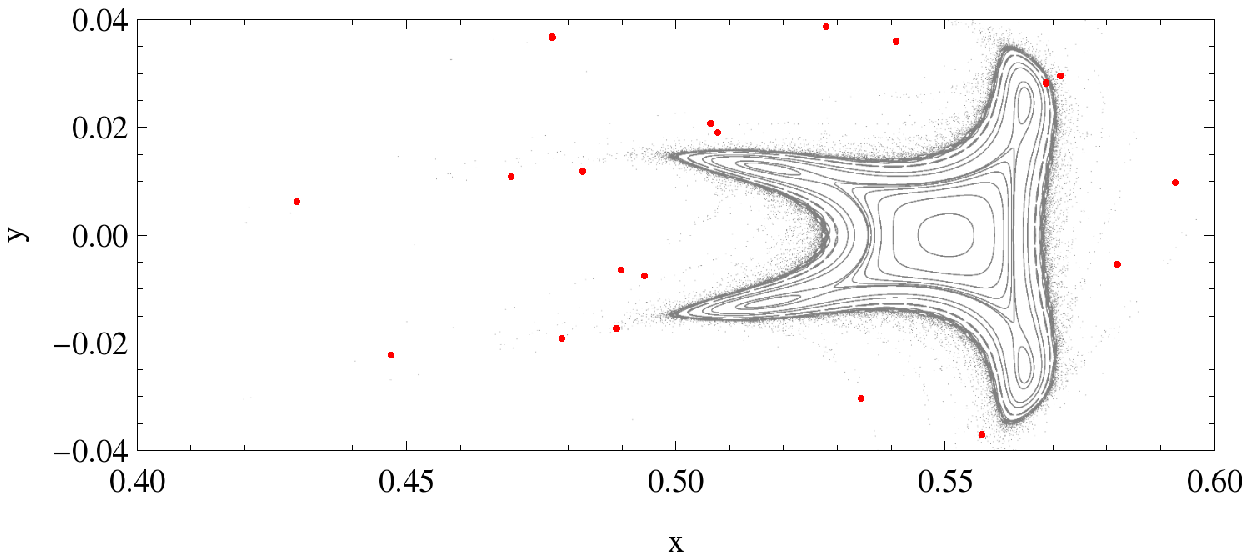}}
\caption{\label{fig3}The surface of section for the a range of energies close to the original energy level $E = -2.152$. The top four figures are at a lower energy and the bottom four figures at a higher energy. The locations of the SOS crossings at the original energy of the initial trajectory are shown in red on each plot. The best potential transfer point is identified as a black point on the $E = -2.1535$ plot}
\end{figure}

The natural chaos-assisted capture method of \cite{Astakhov2003} theorises that orbits similar to this can be moved into the nearby regular region of phase space (at a different energy level) by effects such as gas drag. We propose a method for capturing asteroids that does this through an artificial instantaneous $\Delta v$ applied to the asteroid. Essentially, the chaotic orbit will be `kicked' to a new energy level where the trajectory is now regular. Using the SOS for nearby energy levels we can identify points at which the original trajectory crosses the section nearby to a regular region. The SOSs of nearby energy levels are shown in Figure \ref{fig3}. As noted by \cite{Astakhov2003}, depending on location either an increase or decrease in energy may move the asteroid into a regular KAM region of phase space, and as such the SOS has been plotted for $ -2.154 \leq E \leq -2.150$. It can be seen that the regular region changes fairly significantly with even a small change in energy, and some of the crossings of the SOS of the original asteroid trajectory which are chaotic at $E=-2.152$ are within a regular region of phase space at $E=-2.1535$ and  $E=-2.154$, as well as at $E=-2.1505$ and $E=-2.15$. An instantaneous impulse at one of these points is all that is required to transfer the chaotic orbit to a regular orbit.

Taking the impulse to be perpendicular to the SOS at the moment of transfer allows the $\Delta v$ to be calculated for each potential transfer point. The transfer at $E=-2.1535$ is the best, requiring $\Delta v=64.9\ \mathrm{m\ s^{-1}}$ in the negative $y$ direction. The transfer at $E=-2.1505$ is slightly less favourable, requiring $\Delta v=105.5\ \mathrm{m\ s^{-1}}$ in the positive $y$ direction this time. By comparison, to close the ZVC at the first point through a similar impulse would require a $\Delta v$ of approximately 500 $\mathrm{m\ s^{-1}}$ (and approximately 840 $\mathrm{m\ s^{-1}}$ at the second point). It is clear that the new method leads to significant savings in capture $\Delta v$ cost.

The exact point of the best transfer is marked on Figure \ref{figdv}, and the new trajectory shown in Figure \ref{fig4}. With a relatively low energy requirement the asteroid has been essentially permanently captured in a prograde orbit about the Earth. It has the further advantage of being protected from collision with the Earth as well as escape from the Hill sphere.

The method used here is simple and merely intended to demonstrate the concept. To extend the work to more realistic models a three-dimensional dynamical model of the Earth-Moon-Sun system is essential. An optimization scheme combined with a method to more precisely determine the regions of chaotic and regular trajectories, such as the fast Lyaponuv indicator (FLI) maps used in \cite{Astakhov2003, Astakhov2004, Astakhov2005} would provide a more robust means of detecting the optimal transfer of an asteroid using the chaos-assisted capture method. However, the results presented here demonstrate that even using a simple SOS method to select the transfer point yields very favourable $\Delta v$ requirements.

Finally, the transfer scheme is limited in that it requires the asteroid to be in the chaotic layer, however we note that other authors, such as \cite{Urrutxua2014} have looked at transferring NEOs onto these types of trajectories. The artificial chaos-assisted capture method can therefore be seen as the final capture stage in a more general strategy for transferring asteroids into the Earth neighbourhood.

\section{Conclusion}

The theory of chaos-assisted capture of irregular satellites of the giant planets can be used to develop low-energy capture strategies for certain classes of Near Earth Objects (NEOs) that pass close to a chaotic layer associated with a regular Kolmogorov-Arnold-Moser (KAM) region of phase space. Investigation of the surfaces of section (SOS) within the two-dimensional Hill problem model of the dynamics can be used to identify points along a trajectory at which a low energy instantaneous transfer can be made to move the object into the regular region and onto a long-lived and essentially bound orbit about the Earth. The strategy places the captured object in a regular region of phase space that, in the simple model used, is protected from both collision with the Earth and escape from the Hill sphere. The method of transfer does not require a control strategy, and further neither does the final orbit require any control mechanism in this model. The $\Delta v$ requirements are low, comparable to those needed for temporary capture strategies and much lower than those needed by other permanent capture strategies. Here a $\Delta v$ of order of tens of $\mathrm{m\ s^{-1}}$ is required, compared to the hundreds to thousands of $\mathrm{m\ s^{-1}}$ needed for other capture strategies.

The method is limited in that it is only applicable to asteroids that pass close to a chaotic layer near a regular region of phase space. However, transferring an asteroid artificially into this region of phase space has been addressed by other authors and studies, who consider low energy strategies for transferring asteroids onto longer lived (but still temporary) `sticky' chaotic orbits. Given this, the chaos-assisted capture scheme presented here provides a method for the final stage in the artificial capture of an asteroid.

We also note that the use of an instantaneous impulse to transfer the asteroid to the regular region is likely to be unworkable for any asteroid of significant mass, it is merely a convenient method to illustrate the magnitude of the change in energy required to capture an asteroid using this method. A sustained period of low-thrust impulse would be a more practical solution, and is in fact more in analogy with the original theory of formation of irregular satellites, which suggests that they are  transferred into the regular KAM regions via weak dissipation occurring over a period of time \cite{Astakhov2003}.

The example given in this work is a relatively low energy orbit and here the SOS changes fairly rapidly with energy. For high energy objects the SOS changes slower so the energy requirement and $\Delta v$ will necessarily be larger. However, it worth noting that the true capture $\Delta v$ for such objects can be in the region of tens of thousands of $\mathrm{m\ s^{-1}}$ so the saving in energy is still likely to be significant. It is well known that at higher energies the regular region associated with the prograde orbits no longer exist and the transfer has to be to regions surrounding retrograde orbits. The phase space here contains trajectories that come close to collision with the primary and we note that this provides an interesting opportunity to deflect and capture objects that would otherwise collide with the Earth.

This work is intended to illustrate the potential chaos-assisted capture as a strategy for the artificial capture of asteroids. Extension to a more realistic three dimensional model that includes important perturbations such as the effect of the Moon, as well as full optimization to select the lowest possible energy transfer is necessary. However, even without this the results demonstrate that this is a viable method for the final capture stage of asteroids into the neighbourhood of the Earth.

\section*{Acknowledgments}
This work was funded through the European Research Council, Advanced Investigator Grant 227571 VISIONSPACE. We thank the anonymous referee for helpful comments.

\end{document}